\documentclass[8.5pt,twoside,twocolumn]{article}
\usepackage[super,sort&compress,comma]{natbib} 
\usepackage{mhchem}
\usepackage{times,mathptmx}
\usepackage{secsty}
\usepackage{balance} 

\usepackage{graphicx} 
\usepackage{lastpage}
\usepackage[format=plain,justification=raggedright,singlelinecheck=false,font=small,labelfont=bf,labelsep=space]{caption} 
\usepackage{fancyhdr}
\begin{document}

\thispagestyle{plain}
\fancypagestyle{plain}{
\renewcommand{\headrulewidth}{1pt}}
\renewcommand{\thefootnote}{\fnsymbol{footnote}}
\renewcommand\footnoterule{\vspace*{1pt}%
\hrule width 3.4in height 0.4pt \vspace*{5pt}} 
\setcounter{secnumdepth}{5}

\makeatletter 
\def\subsubsection{\@startsection{subsubsection}{3}{10pt}{-1.25ex plus -1ex minus -.1ex}{0ex plus 0ex}{\normalsize\bf}} 
\def\paragraph{\@startsection{paragraph}{4}{10pt}{-1.25ex plus -1ex minus -.1ex}{0ex plus 0ex}{\normalsize\textit}} 
\renewcommand\@biblabel[1]{#1}            
\renewcommand\@makefntext[1]%
{\noindent\makebox[0pt][r]{\@thefnmark\,}#1}
\makeatother 
\renewcommand{\figurename}{\small{Fig.}~}
\sectionfont{\large}
\subsectionfont{\normalsize} 

\fancyfoot{}
\fancyfoot[RO]{\footnotesize{\sffamily{1--\pageref{LastPage} ~\textbar  \hspace{2pt}\thepage}}}
\fancyfoot[LE]{\footnotesize{\sffamily{\thepage~\textbar\hspace{3.45cm} 1--\pageref{LastPage}}}}
\fancyhead{}
\renewcommand{\headrulewidth}{1pt} 
\renewcommand{\footrulewidth}{1pt}
\setlength{\arrayrulewidth}{1pt}
\setlength{\columnsep}{6.5mm}
\setlength\bibsep{1pt}

\twocolumn[
  \begin{@twocolumnfalse}
\noindent\LARGE{\textbf{Environment Assisted Quantum Transport in Organic Molecules}}
\vspace{0.6cm}

\noindent\large{\textbf{G{\'a}bor Vattay,$^{\ast}$\textit{$^{a}$}  and Istv{\'a}n Csabai,\textit{$^{a}$}}}\vspace{0.5cm}


\vspace{0.6cm}

\noindent \normalsize{One of the new discoveries in quantum biology is the role of Environment Assisted Quantum Transport (ENAQT) in excitonic transport processes. In disordered quantum 
systems transport is most efficient when the environment
just destroys quantum interferences responsible for localization, but the coupling does
not drive the system to fully classical thermal diffusion yet. This poised realm between
the pure quantum and the semi-classical domains has not been considered in other biological transport
processes, such as charge transport through organic molecules. Binding in receptor-ligand 
complexes is assumed to be static as electrons are assumed to be not able to cross the
ligand molecule. We show that ENAQT makes cross ligand transport possible and efficient
between certain atoms opening the way for the reorganization of the charge distribution
on the receptor when the ligand molecule docks. This new effect can potentially  
change our understanding how receptors work. We demonstrate room temperature ENAQT on
the caffeine molecule.}
\vspace{0.5cm}
 \end{@twocolumnfalse}
  ]

\section{Introduction}


\footnotetext{\textit{$^{a}$~Department of Physics of Complex Systems, E{\" o}tv{\" o}s University, P{\'a}zm{\'a}ny P. s. 1/A, H-1117 Budapest, Hungary. Fax: +3613722866; Tel: +36308502614; E-mail: vattay@elte.hu}}




There is overwhelming evidence that quantum coherence plays an essential role in exciton transport in photosynthesis\cite{Nature.10.1038,Nature08811,Panitchayangkoon20072010,PNAS2011}. One of the fundamental
quantum effects there is the Environmentally Assisted Quantum Transport\cite{plenio2008dephasing,LloydGuzik,Mohseni}
(ENAQT). It applies to partially coherent quantum transport in disordered systems. At low temperatures
transport is dominated by quantum walk of excitons over the excitable sites forming a network. While a classical walker 
diffuses away from its initial position like $\sim \sqrt{t}$ in time via taking random turns, a quantum walker takes a quantum superposition of amplitudes of alternative paths. In a strongly disordered system the interference is 
destructive and the walker becomes stuck or 'localized'\cite{Lloyd}. At medium temperatures
coupling to the environment partially destroys quantum interference and the walker is free to move
and diffuse. Then at very high temperatures decoherence becomes very distractive and the exciton gets frozen
due to the Zeno effect\cite{LloydGuzik}. As a result, transport is most efficient at medium temperatures
or at medium level of decoherence and much less efficient at low or high temperatures. Transport efficiency
is the highest and mean transport time is the shortest at medium temperatures. In photosynthetic systems
parameters are such that this optimum is near room temperature (290K).

The conditions for ENAQT look quite generic and one can suspect that it can occur in a wide range
quantum processes in biology. Yet, the presence of ENAQT has only been established in the exciton transport
of light harvesting systems. In this paper we demonstrate that -- indeed -- ENAQT can play a major role
in electron transport through organic molecules at ambient temperatures. Accordingly in certain cases 
our understanding of charge transport in biology needs revision. 

\begin{figure}[h]
\centering
\includegraphics[width=9cm]{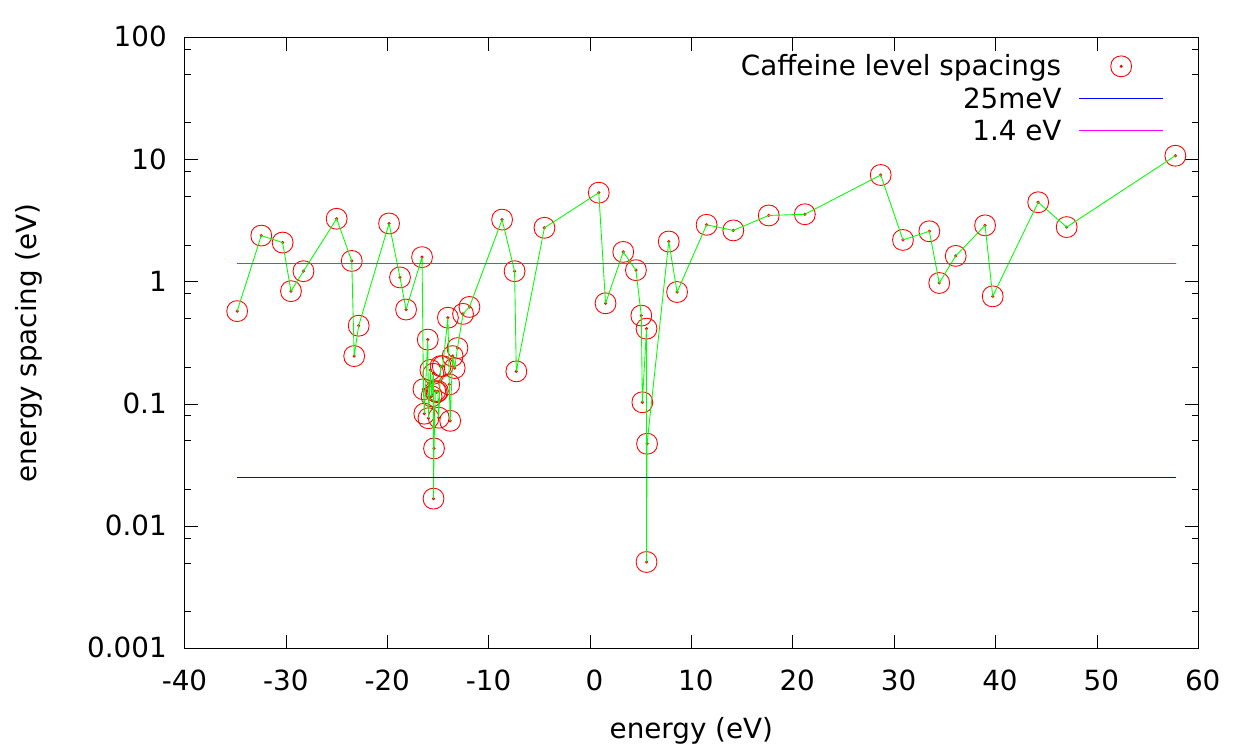}
  \caption{Caffeine electronic energy level spacings $E_{n+1}-E_n$ as a function of the energy $E_n$  are indicated by red circles and connected with green line on a semi-logarithmic plot. Horizontal lines represent the mean level spacing at $\Delta=1.4 eV$ and the room temperature at $k_BT=0.025eV$. Energies are in the $-35.3907eV$~~---~~$+57.7272eV$ range, the HOMO is at $E_{HOMO}=-11.907eV$ and the LUMO is at $E_{LUMO}=-8.6776eV$. Calculation of $N=66$ electronic levels has been carried
out with the Extended H\"uckel method.}
  \label{fgr:fig1}
\end{figure}

Charge transport in biological molecules has been studied extensively with a wide range of quantum chemical
methods\cite{de2012quantum}. These methods cover fully quantum, mixed quantum-classical, semi-classical, 
and fully classical approaches. However, none of these cover the parameter range of the validity of ENAQT.
In general, electronic levels are broadened due to the coupling to the environment. The width of the levels
$\Gamma$ is proportional to the level of decoherence, which is then ultimately determined by the temperature of the environment $\Gamma\sim k_BT$. Quantum description is relevant, when the broadening is much smaller than
the spacing between the consecutive energy levels $E_{n+1}-E_n\gg \Gamma$ and the mean level spacing $\Delta=\left< E_{n+1}-E_n\right>$ is much greater than the temperature $\Delta\gg k_BT$.
The semiclassical and classical or mixed description is relevant in the opposite limit, when $\Delta\ll k_BT$
and many energy levels are involved in the process. The most efficient ENAQT sets in when the environmental temperature is
comparable with the main level spacing $\Delta\approx k_BT$ or at least the level broadening is comparable\cite{celardo2012superradiance} with the spacings of some levels $E_{n+1}-E_n\approx \Gamma\sim k_BT$.
In small and medium sized (less than 500 Da) organic molecules the mean level spacing of electronic levels is on the electron volt scale, wich is about two orders of magnitude larger than the room temperature (about 25 milli-electron volts). So, at first sight we would expect that EQNAT is relevant only at temperatures $T=1eV/k_B\approx 12000K$, which is twice of the temperature of the surface of Sun. However, there is a second possibility: In Fig.~\ref{fgr:fig1} we show the nearest neighbor spacings of the energy levels of caffeine. The mean level spacing is $\Delta=1.4 eV$, which is much larger than the energy scale of the
decoherence $k_BT=0.025eV$. However, there are several level spacings in the spectrum, which are
close to $0.025eV$ indicating that while global ENAQT extending for the entire molecule is not possible, there might be local pockets on the molecule, where it plays an important role. 
Next we continue with showing just that.  

\section{Transport in ligand-receptor complexes}

\begin{figure}[h]
\centering
  \includegraphics[trim = 0mm 30mm 0mm 30mm,clip,width=9cm]{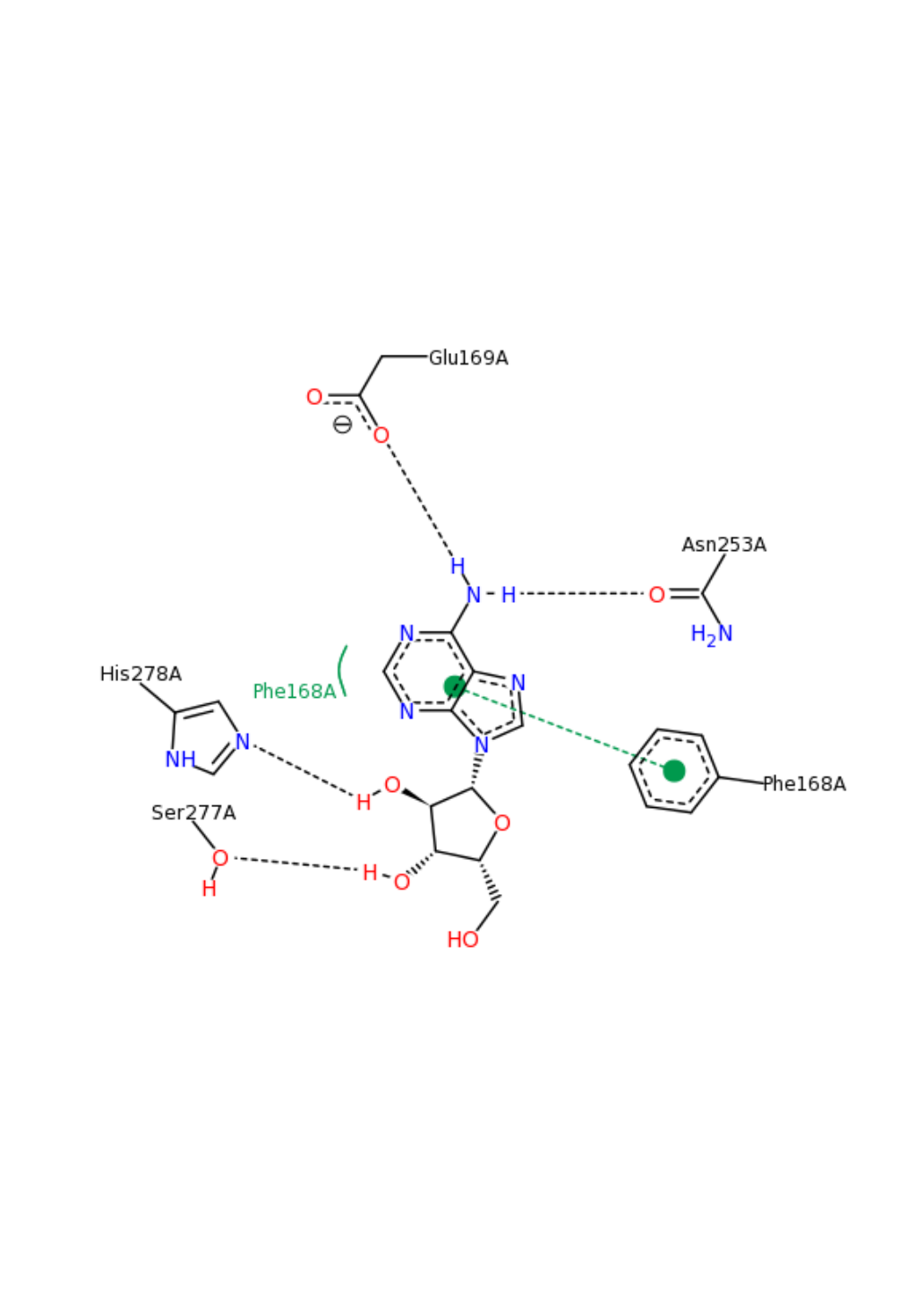}
  \caption{Schematic picture of Adenosine in complex with its A2A receptor. Legend: black dashed lines - hydrogen bonds; green solid lines - hydrophobic interactions; green dashed lines - Pi-Pi, Pi-cation interactions. This figure has been downloaded from RCSB Protein Data Bank and has been created by the
  Pose View software\cite{stierand2010drawing,stierand2007modeling,fricker2004automated,stierand2006molecular}.}
  \label{fgr:fig2}
\end{figure}

One of the most significant building blocks of biochemical processes in the cell
is the docking of signaling molecules in receptors. A typical situation is shown in
Fig.~\ref{fgr:fig2}, where adenosine is docked in the A2A adenosine receptor. 
Besides hydrophobic interactions binding the ligand molecule to the receptor electrostatically, 
there are also hydrogen bonds connecting the two systems. The hydrogen bridge makes possible
the exchange of electrons between the two systems. The receptor consists of amino acids and
these can have very different charging situations. Arginine, histidine and lysine have positive
side chains while glutamic acid and asparatic acid has negative side chains. The rest of the amino
acids is neutral or hydrophobic. An electron from a negatively charged part of the receptor 
protein can jump via the hydrogen bridge to the neighboring atom of the ligand molecule.
If the ligand molecule would be a good conductor the electron could walk trough the molecule and
reach a positively charged part of the receptor trough and other hydrogen bridge. The ligand 
molecule would act as a molecular wire, a lightning rod. As a result, the charge distribution
on the receptor protein would change suddenly upon the docking of the ligand molecule and could
spark sudden motion of the parts of the protein, due to the change of the equilibrium of
the electrostatic forces. Our understanding before ENAQT ruled out such transport of electrons through the ligand molecule. Quantum calculations show that, unless tunneling plays a role in some special cases, ligand molecules act as insulators. Next, we show that ENAQT changes this picture and
provides a mechanism for the easy and effective transport of electrons between certain atoms
in the ligand molecule.

\section{Transport model}

We can develop a model for the description of the electron transfer process of the previous
section. For the description of the molecule only the $n$ electrons 
in the $N$ atomic orbitals are considered. The molecular orbitals $\psi_j$ are linear combinations
of the atomic orbitals 
\begin{eqnarray}
\psi_j=\sum_{i=1}^NC_{jr}\varphi_r&\mbox{and}&j,r=1,...,N,
\end{eqnarray}
where $\varphi_r$ are the valence atomic orbitals $2S,2P_x,2P_y$ and $2P_z$ for carbon atoms and
hetero atoms and $1S$ for hydrogen atoms. The molecular orbitals can be determined from
the overlap matrix $S_{rs}=\langle \varphi_r\mid \varphi_s\rangle$ and Hamiltonian $H_{rs}=\langle \varphi_r\mid H_{eff}\mid\varphi_s\rangle$ matrices via the generalized eigenequation
\begin{equation}
{\bf H}{\bf C}=E{\bf S}{\bf C},
\end{equation}
where ${\bf H}$, ${\bf S}$ and ${\bf C}$ are square matrices containing the elements
of $H_{rs}$, $S_{rs}$ and $C_{jr}$ respectively. The raw molecular orbitals can be
orthonormalized via the L\"owdin transformation\cite{lowdin1950non}. The coefficients
in the L\"owdin basis can be introduced via ${\bf C}={\bf S}^{-1/2}{\bf C'}$ and the
transformed Hamiltonian ${\bf H'}={\bf S}^{-1/2}{\bf H}{\bf S}^{-1/2}$ satisfies the
eigenequation
\begin{equation}
{\bf H'}{\bf C'}=E{\bf C'},
\end{equation}
where the eigenenergies $E$ remain unchanged. Next, we will drop the prime signs and
use this basis as default.

We assume that the electron is coming trough the H-bridge or via other mechanisms
and initially enters one of the atomic orbits of the molecule. 
Initially the $n$ electrons are placed in pairs on the lowest molecular orbits with opposite spins
and occupy about the half of the orbitals. The incoming electron occupies one of the atomic orbitals.
In a pure quantum description the initial wave function is a $n+1$ dimensional Slater determinant of the $n$ molecular orbitals and the single initial electron on the atomic orbital with wave function
$\Psi(t=0)$. This wave function is localized on the atomic orbital and it is
proportional with the initial atomic orbital $\Psi(0)=\alpha\varphi_I$, where $\alpha$ is a 
normalization constant. The initial atomic orbital can be expanded in terms of
all molecular orbitals $\varphi_I=\sum_j \left[{\bf C}^{-1}\right]_{Ij}\psi_j$, where
${\bf C}^{-1}$ is the inverse of the coefficient matrix. The function $\varphi_I$ is not
orthogonal to the $n$ occupied molecular orbitals and only the expansion coefficients 
corresponding to unoccupied molecular orbitals will contribute to the norm of the $n+1$
electron wave function. We can split $\Psi$ into two orthogonal parts
$$\Psi=\Psi^o+\Psi^u,$$
where $\Psi^o$ is spanned over the occupied and $\Psi^u$ over the unoccupied orbitals
of the molecule wit $n$ electrons. The normalization condition of the Slater determinant 
then involves the unoccupied part only $\mid \Psi^u\mid^2=1$,
which yields the following normalization condition
\begin{equation}
\alpha^2\sum_{j\in \mbox{unoccupied}} \left[{\bf C}^{-1}\right]_{Ij}^2=1.
\end{equation}
Consequently the norm $$\mid \Psi\mid^2=\alpha^2=\frac{1}{\sum_{j\in \mbox{unoccupied}} \left[{\bf C}^{-1}\right]_{Ij}^2}$$
is larger than unity. This reflects the fact that in this description the incoming electron 
creates both electronic states in the unoccupied sector and hole states in the occupied sector. 

The time evolution in this model is
very simple. The molecular orbitals are
eigenfunctions of the Hamiltonian and remain unchanged except the stationary phase factors $e^{-iE_jt/\hbar}$. The time evolution of the incoming electron is governed by the Schr\"odinger
equation
\begin{equation}
i\hbar\frac{\partial}{\partial t} \Psi = {\bf H}\Psi.
\end{equation}
The time evolution is then a Slater
determinant again, composed of $e^{-iE_jt/\hbar}\psi_j$ for the occupied orbitals and $\Psi(t)$.

The reduced density matrix of this wave function contains the occupied states and the unoccupied
part $\Psi^u$. In atomic orbital basis
$$\varrho=\mid \Psi^u\left>\right< \Psi^u\mid +\sum_j \mid \psi_j\left>\right< \psi_j\mid.$$ 
This is a very useful expression from the point of view of generalization. The first term represents
the evolution of the incoming electron and the second term represents the rest of the electrons
frozen in the Fermi sea. The incoming electron eventually leaves the system. Just like in case
of light harvesting systems this can be modeled by adding an imaginary sink term to the Hamiltonian
$${\bf H}'={\bf H}-i\kappa \mid \varphi_F\left>\right< \varphi_F\mid,$$
where $\kappa$ is the rate of the leaking out of the electron via a H-bond coupled to
one of the final atomic orbital $\varphi_F$. Once the excess electron is leaked out
the reduced density matrix reduces to the frozen Fermi sea contribution
$$\varrho=\sum_j \mid \psi_j\left>\right< \psi_j\mid.$$
In our non-interacting electron approximation the leaking out of the electron is fully
described by the non-unitary Schr\"odinger equation
\begin{equation}
i\hbar\frac{\partial}{\partial t} \Psi = {\bf H}'\Psi,
\end{equation}
involving now the anti-Hermitian part $-i\hbar\kappa \mid \varphi_F\left>\right< \varphi_F\mid$ describing leaking. The density matrix corresponding to the wave function can be groupped into
four sectors
$$\mid \Psi\left>\right< \Psi\mid=\mid \Psi^u\left>\right< \Psi^u\mid+\mid \Psi^o\left>\right< \Psi^u\mid+\mid \Psi^u\left>\right< \Psi^o\mid+\mid \Psi^o\left>\right< \Psi^o\mid,$$
and only the $(u,u)$ sector has physical meaning. The evolution of density matrix
is described by the von Neumann equation
\begin{equation}
i\hbar\frac{\partial \varrho}{\partial t}  = \left[{\bf H}',\varrho\right]=\left[{\bf H},\varrho\right]-i\left\{{\bf H_1},\varrho\right\},
\end{equation}
where ${\bf H_1}=\hbar\kappa \mid \varphi_F\left>\right< \varphi_F\mid $ is the leaking term
and $\left\{,\right\}$ denotes the anti-commutator. 

We can now take into account the effect of coupling of the electrons to the environment.
This can come from many factors including the phonon vibrations of the molecule and also
very crude interactions with water molecules and other sources of fluctuating electrostatic
forces in the complicated biological environment of a cell. Since all the sources of phase
breaking and dissipation cannot be accounted for we can treat the system statistically and
use the phenomenological approach and add the Lindblad operator to the von Neumann equation
\begin{equation}
i\hbar\frac{\partial \varrho}{\partial t}  = \left[{\bf H}',\varrho\right]+{\cal L}(\varrho).
\label{evolution}
\end{equation}
The most general Lindblad operator\cite{lindblad1976generators} in our case can be written 
in terms of the projection operators of atomic orbitals $A_r=\mid \varphi_r\left>\right< \varphi_r\mid$ as
$${\cal L}=\sum_{rs}L_{rs}\left(2{\bf A_r}\varrho{\bf A_s}-{\bf A_s}{\bf A_r}\varrho-\varrho{\bf A_s}{\bf A_r}\right),$$
where $L_{rs}$ is a positive definite covariance matrix of the noise at different atomic sites. 
We can get $L_{rs}$ from detailed models. The crudest approximation is when we 
assume a completely uncorrelated external noise and set $L_{rs}=\frac{\Gamma}{\hbar}\delta_{rs}$,
where $\Gamma$ is the strength of the decoherence. Its detailed form is model dependent,
but it is in the order of thermal energy and without a detailed knowledge of the system can be set to $\Gamma=k_BT$\cite{cao2009optimization}.

Now the steps of the full calculations can be summarized like this.
First we identify the initial $I$ and final $F$ atomic orbitals, where
the electron enters and exits the system. We initialize the density 
matrix with the initial wave function $\varrho=\alpha^2 \mid \psi_I\left>\right< \psi_I\mid$.
The trace of this density matrix is $Tr{\varrho}=\alpha^2$ and contains both occupied and
unoccupied states. We evolve this density matrix according to (\ref{evolution}). The 
physically relevant reduced density matrix is given by $\varrho^{(u,u)}$ the projection of the density matrix for the unoccupied sector and we use this for the calculation of physical quantities.

\section{Transit time between atoms}

The next step is to calculate physical quantities characterizing the electron
transport between various atoms. In our model we have to characterize the
event of a single electron passing trough the system. Once the electron enters
the system it can leave the system only trough the exit. Unless some symmetry 
consideration forbids it the electron will leave the system with probability $1$,
so we have to find another characteristics. The next quantity is the average time $\tau$ 
an electron needs to get from the entrance to the exit. This depends on two factors
in our model: the details of the molecule and the value of $\kappa$. It is hard to
set the value of $\kappa$ without detailed models or measurements of the system, therefore
we have to find its value from other considerations. Since the typical energy range
in the system is the mean level spacing we can expect that the rate of leaking is
in the order of the time $\kappa\sim \Delta/\hbar$. Once the scale of the escape rate
is set we determine the average time needed from an initial atom to reach the destination 
$\tau_{FI}$ and also the time an electron would need to leave the system if it would be
placed immediately to the final site $\tau_{FF}$. The difference
$\tau_{FI}-\tau_{FF}$ is a better characteristics of the transport as the pure
pure pumping time is transformed out and mainly the inter-atomic travel time enters.

The average transit time can be calculated from the evolution of the reduced density
matrix. The outflowing probability current at the exit site $F$ at time $t$ is given by 
$dP(t)=J_F(t)dt=2dt\kappa \left< \varphi_F\mid \varrho(t)\mid \varphi_F\right>$ and the 
average transit time is the average
\begin{equation}
\tau=2\kappa\hbar\int_0^{\infty}t\left< \varphi_F\mid \varrho(t)\mid \varphi_F\right>dt.
\label{integral}
\end{equation}
This integral can be calculated analytically using the solution of the evolution equation.
Eqation (\ref{evolution}) can be cast into the form
\begin{equation}
\frac{\partial \varrho_{rs}}{\partial t}  = -\frac{i}{\hbar}\sum_{pq}R_{rspq}\varrho_{pq}.
\label{liouville}
\end{equation}
We can re-index this equation and introduce a new index replacing the pairs of indexes
$J=(r,s)$ and $J'=(p,q)$. With the new indexing (\ref{liouville}) reads
\begin{equation}
\frac{\partial \varrho_{J}}{\partial t}  = -\frac{i}{\hbar}\sum_{J'}R_{JJ'}\varrho_{pq}.
\label{liouville2}
\end{equation}
The solution of this is 
$${\bf \varrho}(t)=\exp\left(-\frac{i{\bf R}t}{\hbar}\right){\bf \varrho}(0).$$
Since the initial density matrix has only one nonzero element $J_I=(I,I)$ the
integral in (\ref{integral}) yields the matrix element
\begin{equation}
\tau=-2\kappa \hbar^2 \left[{\bf R^{-2}}\right]_{J_F,J_I},
\end{equation}
where $J_F=(F,F)$. Next, we show the results of the calculation for a molecule with biological relevance.

\section{Caffeine}

\begin{figure}[h]
\centering
  \includegraphics[trim = 0mm 10mm 0mm 10mm,clip,width=9cm]{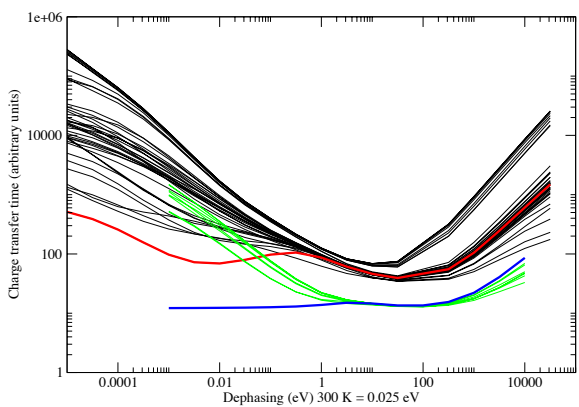}
  \caption{Transit times between atomic orbitals in the caffeine molecule at escape rate $\kappa=\hbar/1eV$.
  Dephasing rate $\Gamma$ in units of $eV$ is shown on the horizontal axis. Curves show transition
  times between the $N=66$ atomic orbitals and the $2S$ orbital of oxygen atom no. 6 in Table~\ref{tbl:atoms}. The blue curve corresponds to the transit time between $2S$ orbital of oxygen no. 6 and itself. The green curves show the transit time between the $2S$ orbital of oxygen no. 6 and the orbits $2P_x,2P_y$ and $2P_z$ of oxygen no. 6. The red curve shows the transition time between the $2S$ orbital of oxygen no. 11 and $2S$ of oxygen no. 6. The black curves show the rest of the transit time curves. }
  \label{fgr:fig3}
\end{figure}

In order to show ENAQT in organic molecules we should pick a 
biologically relevant example which is also computationally feasible. Our choice
is caffeine as it is a sufficiently small molecule with 24 atoms (see in Table~\ref{tbl:atoms}).
Using the Extended H\"uckel method\cite{pople1970approximate}
implemented in the YAeHMOP\cite{yaehmop} software package. There are $N=66$ valence atomic orbitals.
The computational task involves the repeated calculation of the inverse of ${\bf R}$, which is an $(N\times N)\times (N\times N)=N^2\times N^2$ 
matrix, or $4356\times 4356$ in case of caffeine. The rapid growth of the computation time with
$N^2$ is the strongest numerical limitation in the problem. In the calculation the escape rate $\kappa=\hbar/1eV$ has been used. There are 2145 possible pairs of initial and final
atomic orbitals. In Figure~\ref{fgr:fig3} we collected the most interesting results. 
These are the transit times of electrons ending on the $2S$ orbital of oxygen no. 6 in Table~\ref{tbl:atoms}. The blue curve shows the transit time between this orbital and
itself. This is the minimum time $\tau_{FF}$ needed for an electron to leak out from this site.
This value is constant for low dephasing and starts increasing in the classical limit of large dephasing. The green curves show the leaking time from the same atom but the electron is initially
placed on the three $2P$ orbitals of the atom. In this case we can observe a shallow minimum in
the $\Gamma=1-10eV$ range. The black curves represent the transitions between all the atomic orbitals
and the $2S$ orbital of oxygen no. 6, except $2S$ of oxygen no. 11, which is shown in red.
The black and the green curves show the sign of ENAQT and are large both for small and large
$\Gamma$ with a minimum in the middle. The minimum corresponds again to about $1-10eV$ in accordance
with the expectation based on the mean level spacing $1.4eV$, which is 2-3 orders of magnitude larger
than room temperature. The red curve also has a minimum in this range. However, a second shallow minimum is also developed around $0.01eV$ and at room temperature $0.025eV$ the transition time
is still close to this minimum. This second optimum of ENAQT signals fast transport between the
$2S$ orbits of the two oxygen atoms. 

The transition time between the two oxygen atoms is just 10 times as large as the
minimum leaking time set by the scale of $\kappa$. This should be contrasted with 
the quantum limit, where the transition times are 1000 to 100000 times larger than
the minimum leaking time. A pure quantum calculation would find the caffeine molecule
strongly insulating between the atoms and electron transport would be practically impossible.

\begin{table}[h]
\small
  \caption{\ Index and position of atoms in the caffeine molecule }
  \label{tbl:atoms}
  \begin{tabular*}{0.5\textwidth}{@{\extracolsep{\fill}}ccccc}
    \hline
    Index & Atom & x & y & z\\
    \hline
 
   1&    N&  1.047000& -0.000000& -1.312000 \\
   2&    C& -0.208000& -0.000000& -1.790000 \\
   3&    C&  2.176000&  0.000000& -2.246000 \\
   4&    C&  1.285000& -0.001000&  0.016000 \\
   5&    N& -1.276000& -0.000000& -0.971000 \\
   6&    O& -0.384000&  0.000000& -2.993000 \\
   7&    C& -2.629000& -0.000000& -1.533000 \\
   8&    C& -1.098000& -0.000000&  0.402000 \\
   9&    C&  0.193000&  0.005000&  0.911000 \\
  10&    N& -1.934000& -0.000000&  1.444000 \\
  11&    O&  2.428000& -0.000000&  0.437000 \\
  12&    N&  0.068000& -0.000000&  2.286000 \\
  13&    C& -1.251000& -0.000000&  2.560000 \\
  14&    C&  1.161000& -0.000000&  3.261000 \\
  15&    H&  1.800000&  0.001000& -3.269000 \\
  16&    H&  2.783000&  0.890000& -2.082000 \\
  17&    H&  2.783000& -0.889000& -2.083000 \\
  18&    H& -2.570000& -0.000000& -2.622000 \\
  19&    H& -3.162000& -0.890000& -1.198000 \\
  20&    H& -3.162000&  0.889000& -1.198000 \\
  21&    H& -1.679000&  0.000000&  3.552000 \\
  22&    H&  1.432000& -1.028000&  3.503000 \\
  23&    H&  2.024000&  0.513000&  2.839000 \\
  24&    H&  0.839000&  0.513000&  4.167000 \\
 
    \hline
  \end{tabular*}
\end{table}

%
%
 

\section{Conclusions}

We investigated the possibility of ENAQT in organic molecules at room temperature.
We worked out a framework which makes it possible to treat this problem in first
approximation. We showed that global ENAQT is not present in these molecules in
general as the optimal temperature is several orders of magnitude higher than
the room temperature. However, room temperature ENAQT can be observed in the electron
transport between certain atoms in the caffeine molecule. This indicates that 
in ligand-receptor systems the ligand molecule can act as a molecular wire
connecting differently charged areas of the receptor protein and can contribute
to their response for the docking. This can have a significant importance in
understanding signaling and drug action in cells.

The authors thank D. Salahub and S. Kauffman for help, encouragement and numerous 
discussions over the last three years.

\footnotesize{
\bibliography{paper} 

\providecommand*{\mcitethebibliography}{\thebibliography}
\csname @ifundefined\endcsname{endmcitethebibliography}
{\let\endmcitethebibliography\endthebibliography}{}
\begin{mcitethebibliography}{19}
\providecommand*{\natexlab}[1]{#1}
\providecommand*{\mciteSetBstSublistMode}[1]{}
\providecommand*{\mciteSetBstMaxWidthForm}[2]{}
\providecommand*{\mciteBstWouldAddEndPuncttrue}
  {\def\EndOfBibitem{\unskip.}}
\providecommand*{\mciteBstWouldAddEndPunctfalse}
  {\let\EndOfBibitem\relax}
\providecommand*{\mciteSetBstMidEndSepPunct}[3]{}
\providecommand*{\mciteSetBstSublistLabelBeginEnd}[3]{}
\providecommand*{\EndOfBibitem}{}
\mciteSetBstSublistMode{f}
\mciteSetBstMaxWidthForm{subitem}
{(\emph{\alph{mcitesubitemcount}})}
\mciteSetBstSublistLabelBeginEnd{\mcitemaxwidthsubitemform\space}
{\relax}{\relax}

\bibitem[Engel \emph{et~al.}(2007)Engel, Calhoun, Read, Ahn, Man{\v{c}}al,
  Cheng, Blankenship, and Fleming]{Nature.10.1038}
G.~S. Engel, T.~R. Calhoun, E.~L. Read, T.-K. Ahn, T.~Man{\v{c}}al, Y.-C.
  Cheng, R.~E. Blankenship and G.~R. Fleming, \emph{Nature}, 2007,
  \textbf{446}, 782--786\relax
\mciteBstWouldAddEndPuncttrue
\mciteSetBstMidEndSepPunct{\mcitedefaultmidpunct}
{\mcitedefaultendpunct}{\mcitedefaultseppunct}\relax
\EndOfBibitem
\bibitem[Collini \emph{et~al.}(2010)Collini, Wong, Wilk, Curmi, Brumer, and
  Scholes]{Nature08811}
E.~Collini, C.~Y. Wong, K.~E. Wilk, P.~M. Curmi, P.~Brumer and G.~D. Scholes,
  \emph{Nature}, 2010, \textbf{463}, 644--647\relax
\mciteBstWouldAddEndPuncttrue
\mciteSetBstMidEndSepPunct{\mcitedefaultmidpunct}
{\mcitedefaultendpunct}{\mcitedefaultseppunct}\relax
\EndOfBibitem
\bibitem[Panitchayangkoon \emph{et~al.}(2010)Panitchayangkoon, Hayes, Fransted,
  Caram, Harel, Wen, Blankenship, and Engel]{Panitchayangkoon20072010}
G.~Panitchayangkoon, D.~Hayes, K.~A. Fransted, J.~R. Caram, E.~Harel, J.~Wen,
  R.~E. Blankenship and G.~S. Engel, \emph{Proceedings of the National Academy
  of Sciences}, 2010, \textbf{107}, 12766--12770\relax
\mciteBstWouldAddEndPuncttrue
\mciteSetBstMidEndSepPunct{\mcitedefaultmidpunct}
{\mcitedefaultendpunct}{\mcitedefaultseppunct}\relax
\EndOfBibitem
\bibitem[Panitchayangkoon \emph{et~al.}(2011)Panitchayangkoon, Voronine,
  Abramavicius, Caram, Lewis, Mukamel, and Engel]{PNAS2011}
G.~Panitchayangkoon, D.~V. Voronine, D.~Abramavicius, J.~R. Caram, N.~H. Lewis,
  S.~Mukamel and G.~S. Engel, \emph{Proceedings of the National Academy of
  Sciences}, 2011, \textbf{108}, 20908--20912\relax
\mciteBstWouldAddEndPuncttrue
\mciteSetBstMidEndSepPunct{\mcitedefaultmidpunct}
{\mcitedefaultendpunct}{\mcitedefaultseppunct}\relax
\EndOfBibitem
\bibitem[Plenio and Huelga(2008)]{plenio2008dephasing}
M.~B. Plenio and S.~F. Huelga, \emph{New Journal of Physics}, 2008,
  \textbf{10}, 113019\relax
\mciteBstWouldAddEndPuncttrue
\mciteSetBstMidEndSepPunct{\mcitedefaultmidpunct}
{\mcitedefaultendpunct}{\mcitedefaultseppunct}\relax
\EndOfBibitem
\bibitem[Patrick~Rebentrost and Aspuru-Guzik(2009)]{LloydGuzik}
I.~K. S.~L. Patrick~Rebentrost, Masoud~Mohseni and A.~Aspuru-Guzik, \emph{New
  Journal of Physics}, 2009, \textbf{11}, 033003\relax
\mciteBstWouldAddEndPuncttrue
\mciteSetBstMidEndSepPunct{\mcitedefaultmidpunct}
{\mcitedefaultendpunct}{\mcitedefaultseppunct}\relax
\EndOfBibitem
\bibitem[M.~Mohseni and Aspuru-Guzik(2008)]{Mohseni}
S.~L. M.~Mohseni, P.~Rebentrost and A.~Aspuru-Guzik, \emph{J. Chem. Phys.},
  2008, \textbf{129}, 174106\relax
\mciteBstWouldAddEndPuncttrue
\mciteSetBstMidEndSepPunct{\mcitedefaultmidpunct}
{\mcitedefaultendpunct}{\mcitedefaultseppunct}\relax
\EndOfBibitem
\bibitem[Lloyd(2011)]{Lloyd}
S.~Lloyd, \emph{Journal of Physics: Conference Series}, 2011, \textbf{302},
  012037\relax
\mciteBstWouldAddEndPuncttrue
\mciteSetBstMidEndSepPunct{\mcitedefaultmidpunct}
{\mcitedefaultendpunct}{\mcitedefaultseppunct}\relax
\EndOfBibitem
\bibitem[de~la Lande \emph{et~al.}(2012)de~la Lande, Babcock,
  {\v{R}}ez{\'a}{\v{c}}, L{\'e}vy, Sanders, and Salahub]{de2012quantum}
A.~de~la Lande, N.~S. Babcock, J.~{\v{R}}ez{\'a}{\v{c}}, B.~L{\'e}vy, B.~C.
  Sanders and D.~R. Salahub, \emph{Physical Chemistry Chemical Physics}, 2012,
  \textbf{14}, 5902--5918\relax
\mciteBstWouldAddEndPuncttrue
\mciteSetBstMidEndSepPunct{\mcitedefaultmidpunct}
{\mcitedefaultendpunct}{\mcitedefaultseppunct}\relax
\EndOfBibitem
\bibitem[Celardo \emph{et~al.}(2012)Celardo, Borgonovi, Merkli, Tsifrinovich,
  and Berman]{celardo2012superradiance}
G.~L. Celardo, F.~Borgonovi, M.~Merkli, V.~I. Tsifrinovich and G.~P. Berman,
  \emph{The Journal of Physical Chemistry C}, 2012, \textbf{116},
  22105--22111\relax
\mciteBstWouldAddEndPuncttrue
\mciteSetBstMidEndSepPunct{\mcitedefaultmidpunct}
{\mcitedefaultendpunct}{\mcitedefaultseppunct}\relax
\EndOfBibitem
\bibitem[Stierand and Rarey(2010)]{stierand2010drawing}
K.~Stierand and M.~Rarey, \emph{ACS medicinal chemistry letters}, 2010,
  \textbf{1}, 540--545\relax
\mciteBstWouldAddEndPuncttrue
\mciteSetBstMidEndSepPunct{\mcitedefaultmidpunct}
{\mcitedefaultendpunct}{\mcitedefaultseppunct}\relax
\EndOfBibitem
\bibitem[Stierand and Rarey(2007)]{stierand2007modeling}
K.~Stierand and M.~Rarey, \emph{ChemMedChem}, 2007, \textbf{2}, 853--860\relax
\mciteBstWouldAddEndPuncttrue
\mciteSetBstMidEndSepPunct{\mcitedefaultmidpunct}
{\mcitedefaultendpunct}{\mcitedefaultseppunct}\relax
\EndOfBibitem
\bibitem[Fricker \emph{et~al.}(2004)Fricker, Gastreich, and
  Rarey]{fricker2004automated}
P.~C. Fricker, M.~Gastreich and M.~Rarey, \emph{Journal of chemical information
  and computer sciences}, 2004, \textbf{44}, 1065--1078\relax
\mciteBstWouldAddEndPuncttrue
\mciteSetBstMidEndSepPunct{\mcitedefaultmidpunct}
{\mcitedefaultendpunct}{\mcitedefaultseppunct}\relax
\EndOfBibitem
\bibitem[Stierand \emph{et~al.}(2006)Stierand, Maa{\ss}, and
  Rarey]{stierand2006molecular}
K.~Stierand, P.~C. Maa{\ss} and M.~Rarey, \emph{Bioinformatics}, 2006,
  \textbf{22}, 1710--1716\relax
\mciteBstWouldAddEndPuncttrue
\mciteSetBstMidEndSepPunct{\mcitedefaultmidpunct}
{\mcitedefaultendpunct}{\mcitedefaultseppunct}\relax
\EndOfBibitem
\bibitem[L{\"o}wdin(1950)]{lowdin1950non}
P.-O. L{\"o}wdin, \emph{The Journal of Chemical Physics}, 1950, \textbf{18},
  365--375\relax
\mciteBstWouldAddEndPuncttrue
\mciteSetBstMidEndSepPunct{\mcitedefaultmidpunct}
{\mcitedefaultendpunct}{\mcitedefaultseppunct}\relax
\EndOfBibitem
\bibitem[Lindblad(1976)]{lindblad1976generators}
G.~Lindblad, \emph{Communications in Mathematical Physics}, 1976, \textbf{48},
  119--130\relax
\mciteBstWouldAddEndPuncttrue
\mciteSetBstMidEndSepPunct{\mcitedefaultmidpunct}
{\mcitedefaultendpunct}{\mcitedefaultseppunct}\relax
\EndOfBibitem
\bibitem[Cao and Silbey(2009)]{cao2009optimization}
J.~Cao and R.~J. Silbey, \emph{J. Phys. Chem. A}, 2009, \textbf{113},
  13825--13838\relax
\mciteBstWouldAddEndPuncttrue
\mciteSetBstMidEndSepPunct{\mcitedefaultmidpunct}
{\mcitedefaultendpunct}{\mcitedefaultseppunct}\relax
\EndOfBibitem
\bibitem[Pople and Beveridge(1970)]{pople1970approximate}
J.~A. Pople and D.~L. Beveridge, \emph{Approximate molecular orbital theory},
  McGraw-Hill New York, 1970, vol.~30\relax
\mciteBstWouldAddEndPuncttrue
\mciteSetBstMidEndSepPunct{\mcitedefaultmidpunct}
{\mcitedefaultendpunct}{\mcitedefaultseppunct}\relax
\EndOfBibitem
\bibitem[yae()]{yaehmop}
\relax
\mciteBstWouldAddEndPunctfalse
\mciteSetBstMidEndSepPunct{\mcitedefaultmidpunct}
{}{\mcitedefaultseppunct}\relax
\EndOfBibitem
\end{mcitethebibliography}
\bibliographystyle{rsc} 
}

\end{document}